\documentclass[aps,preprint,preprintnumbers,nofootinbib,showpacs]{revtex4}
\usepackage{graphicx,color,amsmath,amsxtra}

\newcommand{\newc}{\newcommand}

\newc{\be}{\begin{equation}}
\newc{\ee}{\end{equation}}
\newc{\beq}{\begin{eqnarray}}
\newc{\eeq}{\end{eqnarray}}

\begin{document}
\title{Radiative neutrino mass generation and dark energy}
\author{K.~Bamba, C.~Q.~Geng and S.~H.~Ho}
\affiliation{Department of Physics, National Tsing Hua University,
Hsinchu, Taiwan 300}
\date{\today}

\begin{abstract}
We study the models with radiative neutrino mass generation and explore the relation
between the neutrino masses and dark energy. In these models, the pseudo-Nambu-Goldston 
bosons (pNGBs) arise at two-loop level via the Majorana neutrino masses. 
In particular, we demonstrate that the potential energy of the pNGB can be the dark energy potential
and the observed value of the equation of state (EoS) parameter of the universe, $i.e.$, 
$w\simeq -1$, can be realized.
\end{abstract}
\maketitle

\section{Introduction}

In the standard model (SM) of particle physics, neutrinos are massless.
However, various experimental searches indicate
that neutrinos have
tiny masses
 ($\leq O(10^{-2})~\mathrm{eV}$) \cite{PDG2006}.
It is a challenging and important problem
 to explain the origin of the small neutrino masses.
Various mechanisms could generate
neutrino
masses~\cite{Mohapatra:1998rq},
in which the one with radiative
neutrino mass generation
without right-handed neutrinos by extending the Higgs
sector \cite{RadNu1,RadNu2,RadNu3} is
particularly interesting because
the neutrino masses
are naturally small. It is clear that without the right-handed
states the active neutrinos can only have Majorana masses.

On the other hand,
recent cosmological observations have confirmed that not only  there existed the
inflationary stage in the early universe, but also  at present
the expansion of the universe is accelerating~\cite{WMAP1, SN1,Frieman:2008sn}.
Although various scenarios for the late-time acceleration in the expansion of
the universe have been proposed,
the cosmic acceleration
mechanism is still not well understood
\cite{Peebles:2002gy, Padmanabhan:2002ji,
Copeland:2006wr, Durrer:2007re, NO-rev}.

In the framework of general relativity,
 the current accelerating universe is
due to
the so-called dark energy (or cosmological constant)
with its density at
the present time is only about $(10^{-3}\mathrm{eV})^4$, which is much smaller
than any known energy scale in particle physics except the neutrino masses.
It is
interesting to note that the energy scale of dark energy coincides with
 the neutrino masses
as discussed in Ref.~\cite{Hill:1988vm}.

Recently, it has been suggested~\cite{Fardon:2003eh,XMZhang,Peccei:2004sz, EV}
that the neutrino masses vary as a function of
a scalar field, called the ``acceleron'', which drives the universe to its
present accelerating phase.
Such neutrinos are referred as mass varying
neutrinos (MaVaNs).
The effects of the MaVaNs
on the anisotropy of the cosmic microwave background (CMB) radiation and
large scale structure (LSS) have been studied in Ref. \cite{CE}.
Several models with
the generation of
the MaVaNs through the see-saw mechanism with right-handed neutrinos
to account for the baryon asymmetry in the universe
have been proposed in Refs.~\cite{Hill:2006hj, Gu:2007ps}.
In these studies, one has to break a
global symmetry spontaneously to get a Nambu-Goldston boson (NGB) and
introduce a soft symmetry breaking term so that
the NGB receives a soft mass via a loop diagram and becomes
a pseudo-Nambu-Goldston boson (pNGB). This pNGB corresponds to
the acceleron field in the scenario proposed in Ref.~\cite{Fardon:2003eh}.
 Some models to explain neutrino masses and dark energy at the TeV scale have been explored
 in Refs.~\cite{Gu:2007gy,Bhatt:2007ah}.
Moreover,
 the Majorana neutrino superfluidity and the stability of the neutrino dark energy
have been  discussed in Ref.~\cite{Bhatt:2008hr}.

In this paper, we consider the generation of the small Majorana neutrino
masses through the radiative mechanism  without
right-handed neutrinos in the framework of the extended
Babu-Zee model~\cite{RadNu2}.
Here, we do not introduce a soft breaking term ``by hand'', but
induce
one from
loop diagrams.
In other words, we break global symmetries spontaneously in the
first place, and then, via loop diagrams, introduce a soft (original) symmetry
breaking term followed by a mass term for the pNGB.
This pNGB also plays a role of the acceleron field.
As a result, we show that
the
small neutrino masses depend on the pNGB and we argue that
the potential energy of the pNGB can be the potential of dark energy.
Furthermore, we demonstrate that the observed value of
the equation of state (EoS) parameter from the
Wilkinson Microwave Anisotropy Probe (WMAP) data on
the anisotropy of the CMB radiation can be realized by following the discussion
in Refs.~\cite{Fardon:2003eh,XMZhang,Peccei:2004sz}.

\section{Pseudo-Nambu-Goldston boson as the acceleron filed}

In the Babu-Zee model~\cite{RadNu2}, it contains only two extra scalar bosons beyond the SM,
$i.e.$, one singly charged scalar ($h^+$) and one doubly charged scalar ($k^{++}$).
In this study, we would extend the Babu-Zee model~\cite{RadNu2}
by considering three singly charged scalars: $h^+_{e\mu}$, $h^+_{e\tau}$ and $h^+_{\mu\tau}$; and three doubly charged scalars: $k^{++}_{e\mu}$, $k^{++}_{e\tau}$ and $k^{++}_{\mu\tau}$,
which  carry different lepton numbers. In addition, we have to introduce
 singlet scalars ($\Phi_{ab}$) to break the lepton number symmetries and induce  non-zero phases.
  The particle contents and  quantum numbers are shown in Table~\ref{table},
\begin{table}[b] \caption{
The particle contents and  quantum numbers, where
$l_{L}$ and $l_{R}$ are the left-handed lepton doublet and right-handed lepton singlet,
 $h^-_{ab}$ and $k^{--}_{ab}$ are singly and doubly charged scalars,
and $a,b=e,\mu$ and $\tau$.
}
 \begin{tabular}[t]{|c|c|c|}\hline
Particles & $SU(2)_L \times U(1)_Y$ & $U(1)_e \times U(1)_{\mu} \times U(1)_{\tau}$ \\ \hline\hline
$l_{La}$ & (2,-1) & $(\delta_{ae},\delta_{a\mu},\delta_{a\tau})$ \\ \hline
$l_{Ra}$ & (1,-2) & $(\delta_{ae},\delta_{a\mu},\delta_{a\tau})$ \\ \hline\hline
$H$ & (2,1) & (0,0,0) \\ \hline
$h^-_{ab}$ & (1,-2) & $(\delta_{ae}+\delta_{be},\delta_{a\mu}+\delta_{b\mu},\delta_{a\tau}+\delta_{b\tau})$\\ \hline
$k^{--}_{ab}$ & (1,-4) & $(\delta_{ae}+\delta_{be},\delta_{a\mu}+\delta_{b\mu},\delta_{a\tau}+\delta_{b\tau})$\\ \hline
$\Phi_{ab}$ & (1,0) &$(\delta_{ae}+\delta_{be},\delta_{a\mu}+\delta_{b\mu},\delta_{a\tau}+\delta_{b\tau})$  \\ \hline
 \end{tabular}
 \label{table}
\end{table}
where $U(1)_e$, $U(1)_{\mu}$ and $U(1)_{\tau}$ represent the electron, muon and tau number symmetries,
 $l_{L}$ and $l_{R}$ are the left-handed lepton doublet and right-handed lepton singlet, $H$ is the
 Higgs doublet in the SM,
and $h^+_{ab}$ and $k^{++}_{ab}$ are singly and doubly charged scalars,
respectively.

The Yukawa couplings between the
singly and doubly charged scalars and  fermions are given by
\beq
\cal{L}_{Y}&=&f_{ab}(l^{Ti}_{aL}C l^j_{bL})\epsilon_{ij}h^+_{ab}+ g_{ab}(l^T_{aR}C l_{bR})k^{++}_{ab}+ \mathrm{h.c.}\,,
\label{Y}
\eeq
where $C$ is the charge conjugation matrix,  $i,j$
and $a,b$ are $SU(2)_L$ and generation indices, respectively.
In our model, it is convenient to expand the Lagrangian in Eq.~(\ref{Y}) as follows:
\beq
\label{Yukawa}
\cal{L}_{Y}&=&2[f_{e\mu}(\bar{\nu^c_e}\mu_L-\bar{\nu^c}_{\mu}e_L)h^+_{e\mu}
                +f_{e\tau}(\bar{\nu^c_e}\tau_L-\bar{\nu^c}_{\tau}e_L)h^+_{e\tau}
           +f_{\mu\tau}(\bar{\nu^c_{\mu}}\tau_L-\bar{\nu^c}_{\tau}\mu_L)h^+_{\mu\tau}
             \nonumber\\
             &&  +g_{e\mu}(\bar{e^c}\mu_R)k^{++}_{e\mu}+ g_{e\tau}(\bar{e^c}\tau_R)k^{++}_{e\tau}+g_{\mu\tau}(\bar{\mu^c}\tau_R)k^{++}_{\mu\tau}]\,\,
            + \mathrm{h.c.}\,,
\eeq
where we have used $f_{ab}=-f_{ba}$, $g_{aa}=0$ and $g_{cd}=g_{dc}\ (c\neq d)$. The Higgs potential can be written as two parts:
\beq
\label{Higgs}
\cal{L}_{\mathrm{1}}&=& \sum_{\xi,\eta}\beta_{\xi}^{\eta}(h^+_{\xi}h^+_{\eta}k^{--}_{\eta}\Phi_{\xi})
+\beta_{ee}^{\mu\tau} (h^+_{e\mu}h^+_{e\tau}k^{--}_{\mu\tau}\Phi_{ee})
+\beta_{\mu\mu}^{e\tau} (h^+_{e\mu}h^+_{\mu\tau}k^{--}_{e\tau}\Phi_{\mu\mu})
\nonumber\\
&&\hspace{0.8cm}+\beta_{\tau\tau}^{e\mu} (h^+_{e\tau}h^+_{\mu\tau}k^{--}_{e\mu}\Phi_{\tau\tau})+ \mathrm{h.c.} \,,
\eeq
where $\xi,\eta=e\mu,e\tau,\mu\tau$ and
\beq
\label{rest}
\cal{L}_{\mathrm{2}}&=& \mu^2(H H^{\dagger})+\lambda (H H^{\dagger})^2
+ \mu_{ij}^2 (h^+_{ij}h^-_{ij})+\lambda_{ij}(h^+_{ij}h^-_{ij})^2
+\tilde{\mu}_{ij}^2 (k^{++}_{ij}k^{--}_{ij})
 \nonumber\\
&&
+\tilde{\lambda}_{ij}(k^{++}_{ij}k^{--}_{ij})^2 +\kappa_{ij} (HH^{\dagger})(h^+_{ij}h^-_{ij})+\tilde{\kappa}_{ij}(HH^{\dagger})(k^{++}_{ij}k^{--}_{ij}) \nonumber\\
&&+ C_{ijlm}(h^+_{ij}h^-_{ij})(k^{++}_{lm}k^{--}_{lm})+ \tilde{C}_{ijlm}(h^+_{lm}h^-_{ij})(k^{++}_{ij}k^{--}_{lm})\,.
\eeq
We remark that  there are enough degrees of freedom to redefine the fields and make all the coefficients in
  Eqs.~(\ref{Yukawa}) and (\ref{Higgs})  real.
  We also note that it is not  necessary to include all three $\Phi_{aa}\ (a=e,\mu,\tau)$
  to get a realistic model. However, to
   have a non-vanishing phase field,
    at least
   one of them is needed. Furthermore, in the model
the lepton symmetries are
spontaneously broken after the singlet scalar fields $\Phi_{ab}$ acquire the vacuum expectation
values (VEVs) $v_{ab}$. For convenience, we parametrize $\Phi_{ab}$ as nonlinear $\sigma$ fields in terms of
the Nambu-Goldstone bosons (NGBs) $\phi_{ab}$ by
\beq
\Phi_{ab}=v_{ab} \exp \left(i\phi_{ab}/v_{ab} \right) \,.
\eeq
For simplicity,  we  assume that all  VEVs
are the same, $i.e.$,
$v_{ab}=v$.
It is clear that there are only three independent NGB states, defined by
\beq
\psi_1=\phi_{e\mu}-\phi_{ee}/2-\phi_{\mu\mu}/2,\ \ \psi_2=\phi_{e\tau}-\phi_{ee}/2-\phi_{\tau\tau}/2,\ \ \psi_3=
\phi_{\mu\tau}-\phi_{\mu\mu}/2-\phi_{\tau\tau}/2\,.
\label{Phases}
\eeq

\section{Neutrino masses}

\begin{figure}[b]
\includegraphics[width=10cm,height=5cm,keepaspectratio=true]{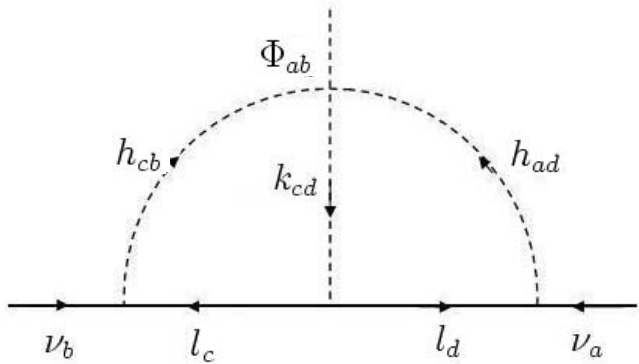}
\caption{Two-loop diagrams to generate neutrino masses.}
\label{massloop}
\end{figure}

As the original Babu-Zee model \cite{RadNu2}, the neutrinos receive Majorana  masses
induced radiatively through  the two-loop diagrams shown
in Fig.~\ref{massloop}.
Consequently, the Majorana neutrino mass term is
found to be~\cite{RadNu2,RadNu3,Babu:2002uu,2loop}
\beq
{\cal L}_{\mathrm{mass}}&=&
-\frac{1}{2}({\cal M}_{\nu})_{ab}(\bar{\nu^c_{L}})_a (\nu_{L})_b+ \mathrm{h.c.}\,,
\nonumber\\
\cal{M}_{\nu} &=& \left(\begin{array}{ccc}
                 M_{ee} & M_{e\mu}e^{i\psi_1/v} & M_{e\tau} e^{i\psi_2/v} \\
                 M_{\mu e}e^{-i\psi_1/v} & M_{\mu\mu} & M_{\mu\tau}e^{i\psi_3/v} \\
                 M_{\tau e} e^{-i\psi_2/v} & M_{\mu\tau}e^{-i\psi_3/v} & M_{\tau\tau}\\
                 \end{array} \right)\,,
\label{massmatrix}
\eeq
with $M_{ab}=M_{ba}$ and
 \beq
M_{ab}&=&\sum_{c,d}8\beta_{ab}^{cd}g_{cd}f_{da}f_{bc}m_cm_dvI_{cd}
\,,
\label{mass}
\eeq
where we have redefined the  neutrinos by
\beq
\nu_{a}\exp(i\phi_{aa}/2v)\rightarrow \nu_{a}\,,
\eeq
$\beta_{ab}^{cd}$ are symmetric under the exchange of $a\leftrightarrow b$ or
$c\leftrightarrow d$
and $I_{cd}$ is the loop integral, given by
\beq
I_{cd}&=&\int \frac{d^4k}{(2\pi)^4} \frac{d^4q}{(2\pi)^4} \frac{1}{(k^2-m_{c}^2)}\frac{1}{(k^2-m_{h_{bd}}^2)}\frac{1}{(q^2-m_{d}^2)}
     \frac{1}{(q^2-m_{h_{ac}}^2)}\frac{1}{((k-q)^2-m_{k_{cd}}^2)}\,,
\label{loopintegral}
\eeq
with $m_{c,d},m_{h_{ac},h_{bd}}$ and $m_{k_{cd}}$ being the corresponding masses of the
charged fermions, singly and doubly charged scalars, respectively.
Here, we have assumed that all couplings of $f_{ab}$ and $g_{ab}$ are real.
We note that all three NGBs become
 pNGBs because the induced neutrino
masses lead to
soft breaking terms into the theory to explicitly break the global
symmetries.
We remark that, unlike the original Babu-Zee model \cite{RadNu2}, there only exist few terms contributing to the neutrino masses because of three global $U(1)$ symmetries.

To estimate $M_{ab}$ in Eq. (\ref{mass}), we take
$m_{c,d}\ll m_{h_{ac},h_{bd}}\sim m_h$ and $m_{c,d}\ll m_{k_{cd}}\sim m_k$.
In these approximations, the loop integral in Eq. (\ref{loopintegral}) can be simplified as~\cite{Babu:2002uu,2loop}
\beq
I_{cd}\simeq\frac{1}{(16\pi^2)^2}\frac{1}{m_h^2}\tilde{I}\left(\frac{m_k^2}{m_h^2}\right)
\,,
\eeq
where the dimensionless function of $\tilde{I}(x)$ is a smooth function from
$\pi^2/3$ decreasing to $\simeq 0.8$ for an interval $10^{-3}<x<10$.
In the Babu-Zee model~\cite{RadNu2} there exist
some tree-level lepton number violating processes. In our
model,  there are no such tree
processes because we introduce more charged scalars couple to different
``pairs'' of leptons.
However, these processes
could be induced at the loop level, which
 will be ignored in our present discussions.
Furthermore, there is no t-channel $k^{++}$ exchange for the muonium-antimuonium
oscillation because there are no  $\bar{e^c_R}e_R k^{++}$ and
$\bar{\mu^c_R}\mu_R k^{++}$ couplings in our model.
Moreover,
there are no lepton number violating radiative decays of $\ell_2\to \ell_1 \gamma \ (\ell_1\neq \ell_2)$
because different pairs of leptons are coupled by different charged scalars.
On the other hand, in our model there are contributions to lepton number conserving
processes such as
 $g-2$ and $\ell_2 \to\ell_1 \nu_2  \bar{\nu}_2$, which give some loose constraints on the couplings of
 $f_{ab}$ and $g_{ab}$ \cite{Babu:2002uu}.
To illustrate the numerical estimation, by choosing that $\beta_{ab}^{cd}\sim 1$,
$f_{ab}\sim 0.1$, $g_{ab}\sim 1$, and $m_h\sim m_k\sim v\sim 1\mathrm{TeV}$, all neutrino mass elements
in Eq. (\ref{mass}) are
found to be $\leq O(10^{-2}\mathrm{eV})$. We note that since there are too many free parameters in the model, it is always possible to obtain the realistic neutrino mixings.
Finally, we remark that the three singlet fields of $\Psi_{aa}\ (a=e,\mu,\tau)$ can be reduced to one without
altering the feature of the model.
%

\section{Dark energy}

Similar to Ref.~\cite{Barbieri:2005gj} (see also~\cite{Gu:2007ps}),
we can write down the induced potential for pNGBs as follows:
\beq
V(\psi_1,\psi_2,\psi_3)\sim\frac{1}{32\pi^2} \mathrm{Tr}
\left[ \cal{M_{\nu}}\cal{M_{\nu}}^{\dagger}\cal{M_{\nu}}\cal{M_{\nu}}^{\dagger}
              \mathrm{ln\frac{\Lambda^2}{\cal{M_{\nu}}\cal{M_{\nu}}^{\dagger}}} \right]\,,
\label{potential}
\eeq
where $\Lambda$ is the ultraviolet cutoff. Expanding Eq.~(\ref{potential}),
we obtain
\beq
V(\psi)\sim \left[
\frac{1}{4\pi^2}\rho \cos \left( \frac{\psi}{v} \right)
+\emph{O}(M_{ab}^4)
\right]
\ln \frac{\Lambda^2}{\Lambda^2_{\rho}}\,,
\label{potential-1}
\eeq
where $\rho  \equiv (M_{ee}+M_{\mu\mu}+M_{\tau\tau})(M_{e\mu}M_{\mu\tau}M_{e\tau})$, $\Lambda^2_{\rho}\equiv M^2_{ee}+M^2_{\mu\mu}+M^2_{\tau\tau}+2M^2_{e\mu}+2M^2_{\mu\tau}+2M^2_{e\tau}$ and $\psi \equiv \psi_1-\psi_2+\psi_3$. Note that in our minimal model, $M_{ee}$ and
$M_{\mu\mu}$ are zero.
We shall concentrate on
the  field of $\psi$, which is the linear combination of the pNGBs. We will demonstrate that in our model, 
the field $\psi$ plays 
a role of the 
acceleron field with the potential $V\sim\emph{O} (m_{\nu}^4)$ and
the effective mass $m_{\psi}^2 \sim \emph{O}(m_{\nu}^4/v^2)$.
To do this,
we  first examine
the case in which the energy density in the dark sector
$\rho_{\mathrm{dark}}$ is
made of
the densities of neutrinos ($\rho_{\nu}$)
and  dark energy ($\rho_\mathrm{DE}$), given by
\begin{eqnarray}
\rho_{\mathrm{dark}} = \rho_{\nu} + \rho_\mathrm{DE}\,,
\label{eq:E1}
\end{eqnarray}
where the dark energy density is assumed to be a function of
neutrino masses ($m_\nu$ ), $i.e.$, $\rho_{DE} = \rho_\mathrm{DE} (m_\nu)$.
 At the present time, because neutrinos are nonrelativistic,
 $\rho_{\nu} = m_\nu n_\nu$, where
$n_\nu$ is the total number density of neutrinos and antineutrinos. Hence,
from Eq.~(\ref{eq:E1}) we get
\begin{eqnarray}
\rho_{\mathrm{dark}} = m_\nu n_\nu + \rho_\mathrm{DE} (m_\nu)\,.
\label{eq:E2}
\end{eqnarray}
Here, we have
concentrated on
$\rho_{\mathrm{dark}}$ being stationary with
respect to the variation of the neutrino masses,
which implies that
\begin{eqnarray}
\frac{\partial \rho_{\mathrm{dark}}}{\partial m_\nu}
= n_\nu + \frac{\partial \rho_\mathrm{DE} (m_\nu)}{\partial m_\nu}
= 0\,.
\label{eq:E3}
\end{eqnarray}

By defining the equation of state (EoS) parameter $w$ as
$w=p/\rho$, where
$p$ is the total pressure of the dark sector of the universe and $\rho$ is the
total energy density of it,
we find~\cite{Fardon:2003eh, Peccei:2004sz}
\begin{eqnarray}
w+1 \simeq \frac{m_\nu n_\nu}{\rho_{\mathrm{dark}}}
    \simeq -\frac{m_\nu}{\rho_{\mathrm{dark}}}
\frac{\partial V_\mathrm{DE}(m_\nu)}{\partial m_\nu}\,,
\label{eq:E4}
\end{eqnarray}
where in deriving the second approximate equality we have used
Eq.~(\ref{eq:E3}) and $\rho_\mathrm{DE} (m_\nu) \simeq V_\mathrm{DE}(m_\nu)$
and $V_\mathrm{DE}(m_\nu)$ is the dark energy potential.
We note that for
the  approximation in Eq. (\ref{eq:E4}),
we have neglected the contribution of any kinetic terms to
the dark energy density\footnote{The accuracy of this approximation is
shown in Ref.~\cite{Fardon:2003eh}.}.

We now discuss
the case in which the neutrino masses depend on some
scalar field $\mathcal{A}$, called the acceleron,
$m_\nu = m_\nu (\mathcal{A})$.
 From Eq.~(\ref{eq:E4}) we obtain
\begin{eqnarray}
w+1 \simeq -\frac{m_\nu}{\rho_{\mathrm{dark}}}
\frac{\partial V_\mathrm{DE}(m_\nu)}{\partial \mathcal{A}}
\frac{1}{\partial m_\nu/ \left( \partial  \mathcal{A} \right)}\,.
\label{eq:E5}
\end{eqnarray}

According to the five-year WMAP data on the anisotropy of the CMB
radiation~\cite{Komatsu:2008hk}, the observed value of $w$ is
$w \simeq -1$.
To have $w \simeq -1$,
it follows from Eq.~(\ref{eq:E5}) that
the potential for $\mathcal{A}$ has to be very flat, $i.e.$,
$\partial V_\mathrm{DE}(m_\nu)/ \left(\partial \mathcal{A}\right) \simeq  0$,
and/or the dependence of $m_\nu$ on $\mathcal{A}$ has to be very steep, $i.e.$,
$\partial m_\nu/ \left( \partial  \mathcal{A} \right) \gg 1$.

Hence,
in our model
the pNGB $\psi$ corresponds to the acceleron field $\mathcal{A}$.
 From Eq.~(\ref{potential-1}), we find that
the dark energy potential is given by
\begin{eqnarray}
V_\mathrm{DE} = V(\psi) \sim m_{\nu}^4
\cos \left( \frac{\psi}{v} \right)\,.
\label{eq:E6}
\end{eqnarray}
For $|\psi/v| \ll 1$,
which can be satisfied if the scale $v$ is
sufficiently large,
we find that
\beq
|\partial V(\psi)/ \left(\partial \psi \right)|
= \left( m_{\nu}^4/v \right) \sin \left( \psi/v \right) \simeq 0\,.
\label{eq:E7}
\eeq
Thus, it follows from
Eqs.~(\ref{eq:E5}) and (\ref{eq:E7}) that
in this model the observed relation $w \simeq -1$ can be realized.

\section{Summary}

In summary, we have considered the generation of the small neutrino
mass through the radiative mechanism in the
extended Babu-Zee models.
 We have shown that
the generated small neutrino masses depend on a pNGB, which can
play a role of the acceleron field and
the potential energy of the pNGB can be the dark energy potential.
In particular, we have demonstrated that the observed value of
the EoS parameter from WMAP can be realized.

\begin{acknowledgments}
We would like to thank Prof. T.~C.~Yuan for discussions. This work is supported in part by
the National Science Council of
R.O.C. under Grant \#:
NSC-95-2112-M-007-059-MY3 and National Tsing Hua University under Grant \#: 97N2309F1.
\end{acknowledgments}


\end{document}